\documentclass{b98proc}

\include{psfig}

\def\Journal#1#2#3#4{{#1} {\bf #2}, #3 (#4)}

\def\NPA{{\em Nucl. Phys.} A}

\def\PLB{{\em Phys. Lett.} B}

\def\PRL{\em Phys. Rev. Lett.}

\def\PRD{{\em Phys. Rev.} D}

\begin{document}

\title{Heavy Baryons - Different Facets of Experimental Results}
\author{Stephan Paul}
\address{Physik Department E18\\Technische Universit\"at M\"unchen\\85747 Garching b. M\"unchen}%



\maketitle
\section{Introduction}
In this article we will discuss the various facets of heavy
baryons considering the life cycle of such objects. Although the
physics involved is very different (it spans from perturbative and
non-perturbative QCD describing heavy baryon production (birth)
over quark-models and Heavy Quark Effective Theory (HQET) modeling
the spectroscopy (life) to Heavy Quark Expansion used to describe
their lifetimes and decay asymmetries (death). At the end we will
give an outlook into the future, describing further studies worth
being performed on this specimen.

The field of light baryons, made up of u and d-quarks has been
subject of studies for many decades and has led us to a fist
understanding of baryon structure and the underlying forces. One
might thus ask the question why the study of heavy baryons
{\footnote {Although heavy baryons usually are referred to as
being made from at least a c- or b-quark, we will also include some
strange baryons in this paper, as no other talk on this conference
dealt with them.}} is still of any interest, in particular as this
specimen is difficult to come by, the study of their properties is
thus an experimental challenge. The common answer of the theorists
is clear: the heavy quark constitutes a static colour source which
reduces the 3-body problem to an effective 2-body one. Static
colour sources can be dealt with 'easily' and their transitions
into other (almost) static colour sources can be described
theoretically much easier than in the relativistic case.

\par
The study of such objects also allows to probe existing models in
a different mass regime. In addition, the quasi decoupling of the
heavy quark from the light quarks (in the spin degrees of freedom)
leads to a new arrangement of relevant quantum numbers and a
reordering of states concerning their mass splitting. This has
practical consequences as heavy baryons with large excitation
energy turn out to have rather narrow decay width and are 'easy'
to observe as compared to their light colleagues.

\par
One might still ask the question: 'who cares ?'. Lets briefly give some
arguments why the study of the different aspects of life is still
interesting and gives heavy baryons a particular place in hadron physics.

\begin{itemize}
  \item {Production} - distinguishes {\it hard} and {\it soft}
  processes ({\it QCD} and {\it fragmentation})
  \item {Mass splitting} - tests mass dependence of the effective
  interaction\\
  - separates degrees of freedom ({\it light} versus {\it heavy}).
  \item {Decays} - separates quark-{\it decay} and quark-{\it correlations} in
  hadrons
\end{itemize}

\section{Birth - Heavy Quark Production}
The production of heavy baryons is usually treated as a two-step
process, the production of a heavy quark pair and its subsequent
hadronization. While the heavy quark production is a process which
can be described in perturbative field theories (QCD for
hadro-production and QED or SU(2)xU(1) for electro-weak production) the
fragmentation into hadrons is subject to models. The heavy quarks must find
suitable partners (typically light quarks) to form observable
hadrons which requires their existence in the small part of phase
space allowing a fusion of all future constituents. These other
quarks can either be created in the fragmentation process (the
successive breakings of the colour flux tube stretched between the
heavy quark pair) or could be picked from a premordially existing
system (the remnants of the hadrons participating in the initial
interaction). While the first process has been studied to great
detail in $e^+e^-$ collisions and is now modeled with good
precision based on Monte-Carlo techniques, the latter process is
still not understood.

\subsection{Total cross sections in hadro production}
We shall thus start with an overview on hadro-production of heavy
baryons.
Fig.1
shows the cross section of
heavy quark pair production as a function of the effective beam
energy in fixed target experiments with incoming baryons. The
figure depicts two features.
\begin{itemize}
  \item The curves denote the result of QCD-calculations\cite{ridolfi}. In
  these calculations factorization is assumed to separate {\it soft} and
  {\it hard} processes calculated to second order. The different
  curves denote the results under variation of different
  parameters, the quark mass and the factorization scale. While
  the effective quark mass is a principle unknown the strong
  dependence on the factorization scale denotes the effective
  problems of the calculations.
  In general the cross section is expected to rise with beam
  energy.

  \begin{minipage}[h]{5cm}
  \item The large spread in experimental results over a wide range
  of energies shows the conceptual problem of extracting the
  total charm cross section from data. Experiments usually detect
  a limited number of different charmed hadrons (baryons and mesons)
  in only part of the phase space. Thus, extrapolation to the rest
  of the phase space and to unobserved species has to be done.
  Since $q$-$\overline{q}$-correlations (and the
  respective hadron-correlations which are governed by leading particle effects)
  are not known, charm counting
  leads to systematic uncertainties.
  \end{minipage}
\end{itemize}

\vspace{-8.3cm}

\hspace{6cm}

\begin{figure}[h]
\hspace{6cm}
\psfig{figure=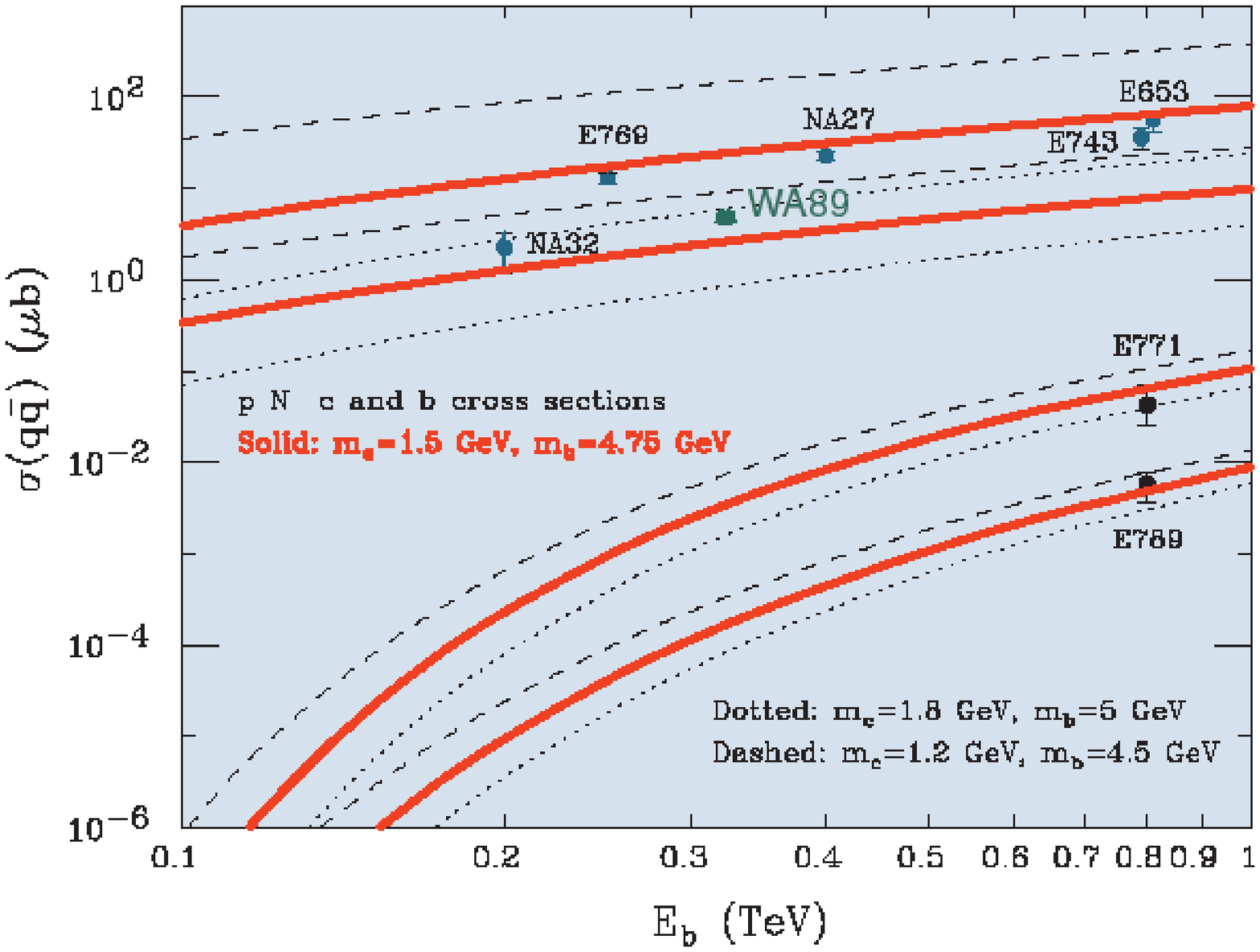,height=5.5cm,width=6.cm}
\label{hadroproduction}

\hspace{6.cm}
\begin{minipage}[htb]{6cm}
\caption{Hadroproduction cross sections for heavy flavours. The
pair of lines denote different results depending on the quark
masses and factorisation scales.}
\end{minipage}
\end{figure}

\vspace{0.5cm}

\subsection{Heavy quark hadronization in hadro-production}
\begin{figure}[h]
\psfig{figure=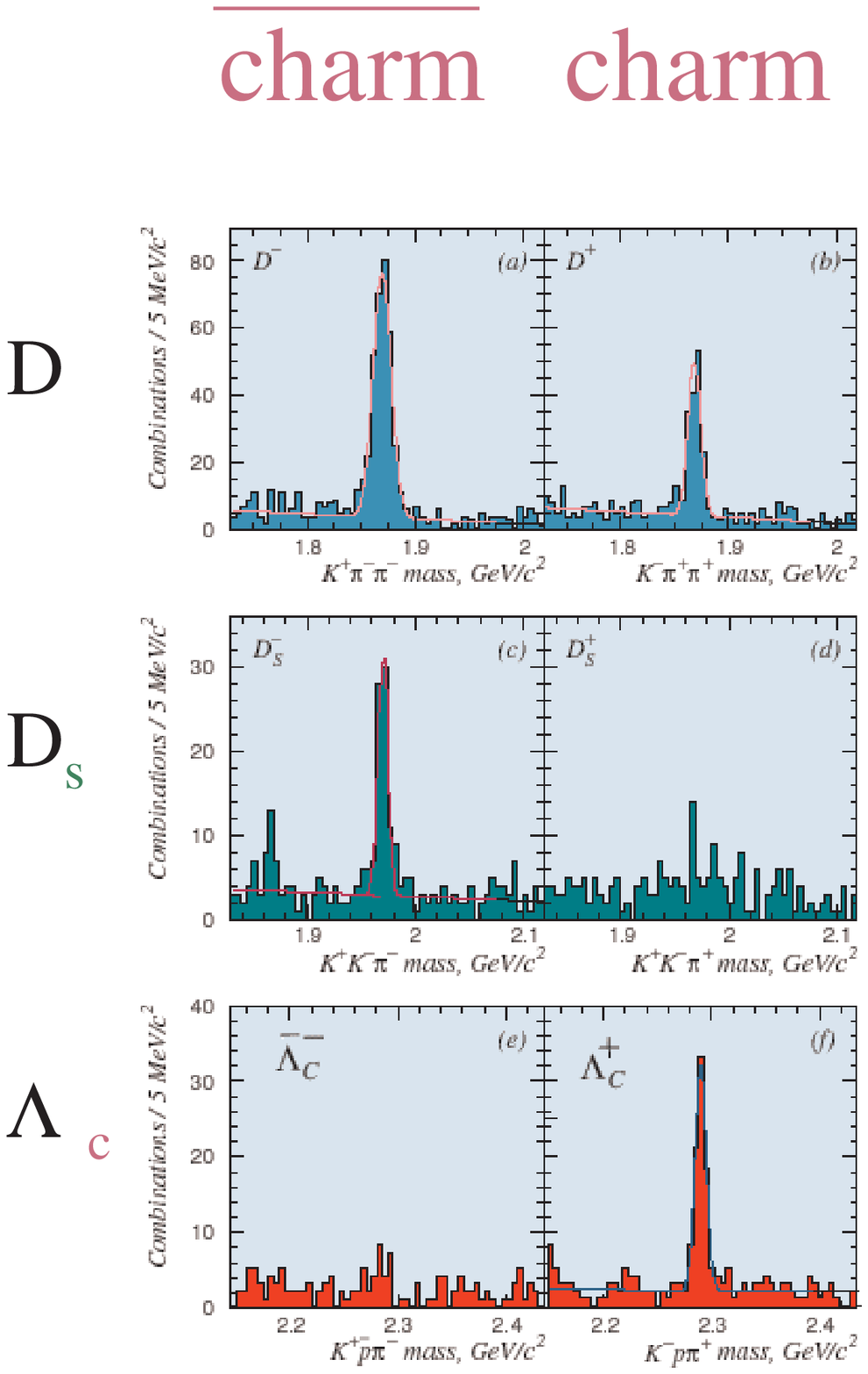,height=7cm,width=6cm}
\psfig{figure=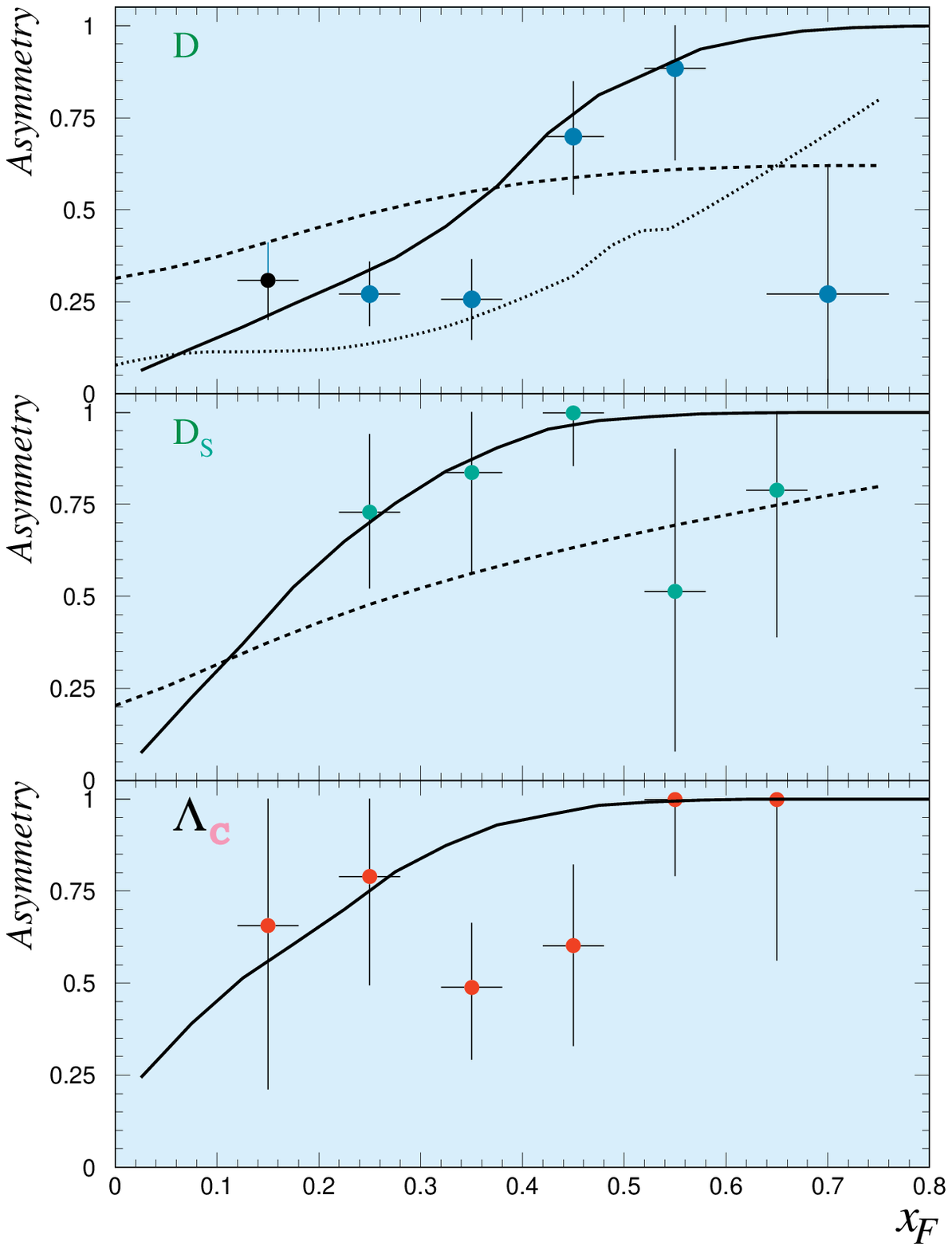,height=7cm,width=6cm}

\hspace{6.5cm}
\caption{Charm production asymmetries observed in a
330 GeV/c hyperon beam at CERN. Left and right histograms show the 
invariant mass distributions of final states for D-, D$_s$-mesons,
$\Lambda_c$-baryon and their anti-particles, respectively}
\label{wa89_1}
\end{figure}

Fig.\ref{wa89_1} shows the resultant signals observed in charm
production by high energy $\Sigma^-$-hyperons of 330 GeV/c in the
WA89 experiment at CERN \cite{WA89_charm}. The hyperon carries strangeness and
baryon-number and both can be found in the kinematical projectile
region ($x_F\geq$ 0.3), where the experiment has large acceptance.
We see a dominant asymmetry in the baryon over anti-baryon
production ($\Lambda_c$ over $\overline{\Lambda_c}$) and an
efficient association of the $\overline{c}$-quark with a strange
quark in the meson sector. The typical preference of hadronization
into mesons is not observed and the asymmetries in the meson
sector indicate an 'eating up' of the c($\overline{c}$)-quarks by the quark
remnants of the projectile.

\hspace{6cm}

\par
This feature is also confirmed by first data from the hyperon beam
experiment E781 at FNAL using a 600 GeV/c beam \cite{SELEX}. Enough statistics
could be accumulated to determine with good accuracy the
$x_F$-dependence of $\Lambda_c$ production for different
projectiles.
Fig.3 shows observed signals (preliminary
data) for $\Lambda_c$ and D-mesons (and their anti-particles) for
p, $\pi$ and $\Sigma^-$ beams. Again, large baryon asymmetries are
observed for baryon beams (much less for $\pi$-beam), much smaller
and reversed asymmetries for D-mesons.

\subsection{Photo-production of heavy baryons}
While charm production constitutes only about 1/1000 of the total
cross section in hadron beams this ratio is much more favorable
for photon beams (1/100). This property is the basis of the very
large data sample on charmed hadrons obtained by the latest photo
production experiment at FNAL, E831.
A total sample of about 10k $\Lambda_c$ in the final
state pk$\pi$ may be expected from this experiment\cite{FOCUS} with an
excellent signal/background ratio of almost 10. Combining
$\Lambda_c$, selected with less stringent criteria, with slow
pions in the same event shows signals for the well known $\Sigma_c$
in all three charged states as well as the excited $\Lambda_c^*$,
when combined with two pions, using only about 10\% of their data.
Although baryon production
constitutes only about 1/10 of charmed hadrons in photon induced
reactions (in the forward hemisphere) these data will constitute
the largest data set in this field for the near future.
\par
\noindent
\begin{minipage}[h]{5cm}
In $e^+e^-$-colliders heavy quarks are produced via the coupling
of the photon to the charge of the quark, resulting in very large
partial cross sections
($\sigma_{c\overline{c}}/\sigma_{tot}\approx 0.3$ and
$\sigma_{b\overline{b}}/\sigma_{tot}\approx 0.1$ at
$\sqrt{s}$=10.4 GeV). CLEOII, running at the CESR storage ring at
Cornell cannot produce beauty baryons owing to the low energy.
Charmed baryons are either produced by fragmentation of the
c-quarks produced (leading to a hard momentum distribution) or by
decays of beauty mesons (resulting in a soft momentum
distribution).
The latter sample does not play a role in most of
their analysis owing to the condition that the baryon should carry
at least 50\% of the beam momentum which is applied to clean the
data samples.
\end{minipage}

\vspace{-10.5cm}

\begin{figure}[h]
\hspace{5.5cm}
\psfig{figure=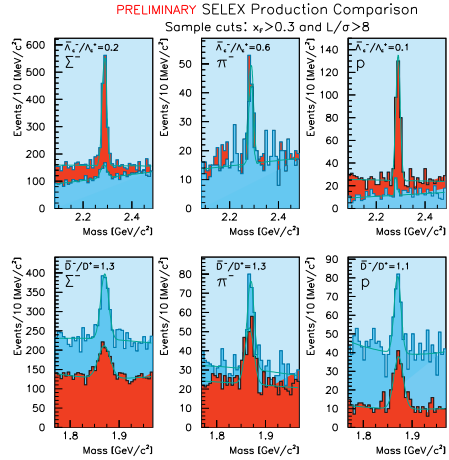,height=6.5cm,width=6.5cm}

\hspace{6cm}
\begin{minipage}[h]{5.5cm}
\caption{Charm production asymmetries observed in a
600 GeV/c negative beam at FNAL. The pictures show the invariant
mass plots for leading and non-leading charmed hadrons for different projectiles. Data are preliminary.}
\end{minipage}

\label{E781}
\end{figure}

\vspace{-0.5cm}

At very high energies $\sqrt{s}$ of about 90 GeV (at LEP), heavy
quark production is governed by the electro-weak coupling to the
quarks ($\sigma_{c\overline{c}}/\sigma_{tot}\approx 0.13$ and
$\sigma_{b\overline{b}}/\sigma_{tot}\approx 0.16$). Again, baryons
constitute about 10\% of the resulting heavy hadrons \cite{aleph_baryon_counting}.

\section{Life of heavy baryons - Spectroscopy}
The mass of a heavy baryon can be viewed as a sum of quark rest masses
and interaction energy. In a constituent quark model\cite{Isgur} depicted in
fig.4
these terms are:\\

\hspace{-1cm}
\begin{minipage}[h]{7.5cm}
\begin{itemize}
  \item effective masses of constituent quarks
  \item kinetic energy
  \item two-body interaction:
    \begin{itemize}
      \item potential energy (confinement)\\
      \item spin-spin interaction ($\mu_q\sim$ 1/$m_q$)\\
        $\Delta$m($\Lambda-\Sigma$) increases with $m_Q$\\
        $\Delta$m($\Sigma^*-\Sigma$) decreases with $m_Q$
     \item spin-orbit interaction (tensor force)
    \end{itemize}
\end{itemize}
\end{minipage}

\vspace{-7cm}

\hspace{0cm}

\hspace{8.5cm}

\begin{figure}[ht]
\hspace{8.5cm}
\psfig{figure=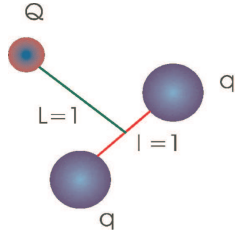,height=3.cm,width=3.cm}

\hspace{8.5cm}
\begin{minipage}[h]{3cm}
\caption{Separation of quark degrees of freedom in a heavy baryon.}
\end{minipage}
\label{isgur_picture}
\end{figure}

\vspace{.95cm}

As mentioned in the introduction the distinct heavy mass of one
constituent inside a baryon leads to a separation of {\it
heavy/light} degrees of freedom. This may become clear in the
following example:
\newline
Consider the case for unit orbital momentum. We can have l=1 of
the light quark pair. $J^p$ of the light quark pair results from a
coupling of its total spin S with l=1 ($0^+,1^+ \otimes 1^-$). To
this, the heavy quark ($J^p$=$1/2^+$) couples (although weakly due
to $H_{int}\sim 1/m_Q$ leading to almost digenerate doublets)
resulting in 7
different states, for $\Lambda$ and $\Sigma$, respectively. The
resulting states will separate in energy from the case of L=1,
orbital momentum between the light quark pair and the heavy quark
(for $\Lambda$ L=1 states have lower energy than l=1 states since the attractive
spin-spin force in the light diquark requires large wave function
overlap, thus l=0).
Since the configurations become distinctly different (since l and
L become different quantum numbers) the decay path of l=1 states
is altered, thus altering also their decay widths (see section 3.2).

\subsection{Measurement of the mean square charge radius of the
$\Sigma^-$}
Before we probe baryon structure using spectroscopic
data, we shall first consider the classical approach of electron
scattering applied to simple strange baryons. This year has
seen the first determination of the mean square charge radius of
the $\Sigma^-$-baryon, using electron scattering in inverse
kinematics. The technique of scattering beams of unstable
particles from electrons off a target nucleon had already been
employed for the $\pi$ and kaon. In the two hyperon beam
experiments at CERN and FNAL heavy targets have been used for the
first time. The technique was developed by WA89. Events were
identified and analyzed using a kinematic fit to all observables
(momentum measurement of incoming and outgoing hyperon,
determination of electron emission angle and its momentum).

\noindent
\begin{minipage}[h]{5cm}
The $Q^2$ dependence of the scattering cross section in this first
measurement is depicted in fig.5
The
resulting fit to the slope of the form factor at $Q^2$=0 gives
\(<$ r$^2$\(>$$_{\Sigma}$=0.91$\pm$0.32$\pm$0.4 fm$^3$ with the first error
reflecting the statistical uncertainty and the second one the
systematic one\cite{WA89_sigel}. E781 has obtained larger samples and owing to
their larger beam momentum also obtained a larger accessible
$Q^2$-range. Their preliminary result is
\(<$ r$^2$\(>$$_{\Sigma}$=0.60$\pm$0.08(stat.)$\pm$0.08(syst.) fm$^3$(see also
contribution of I.Eschrich to this conference\cite{SELEX_sigel}).
\end{minipage}

\vspace{-8.cm}
\hspace{6cm}

\begin{figure}[ht]
\hspace{5.8cm}
\psfig{figure=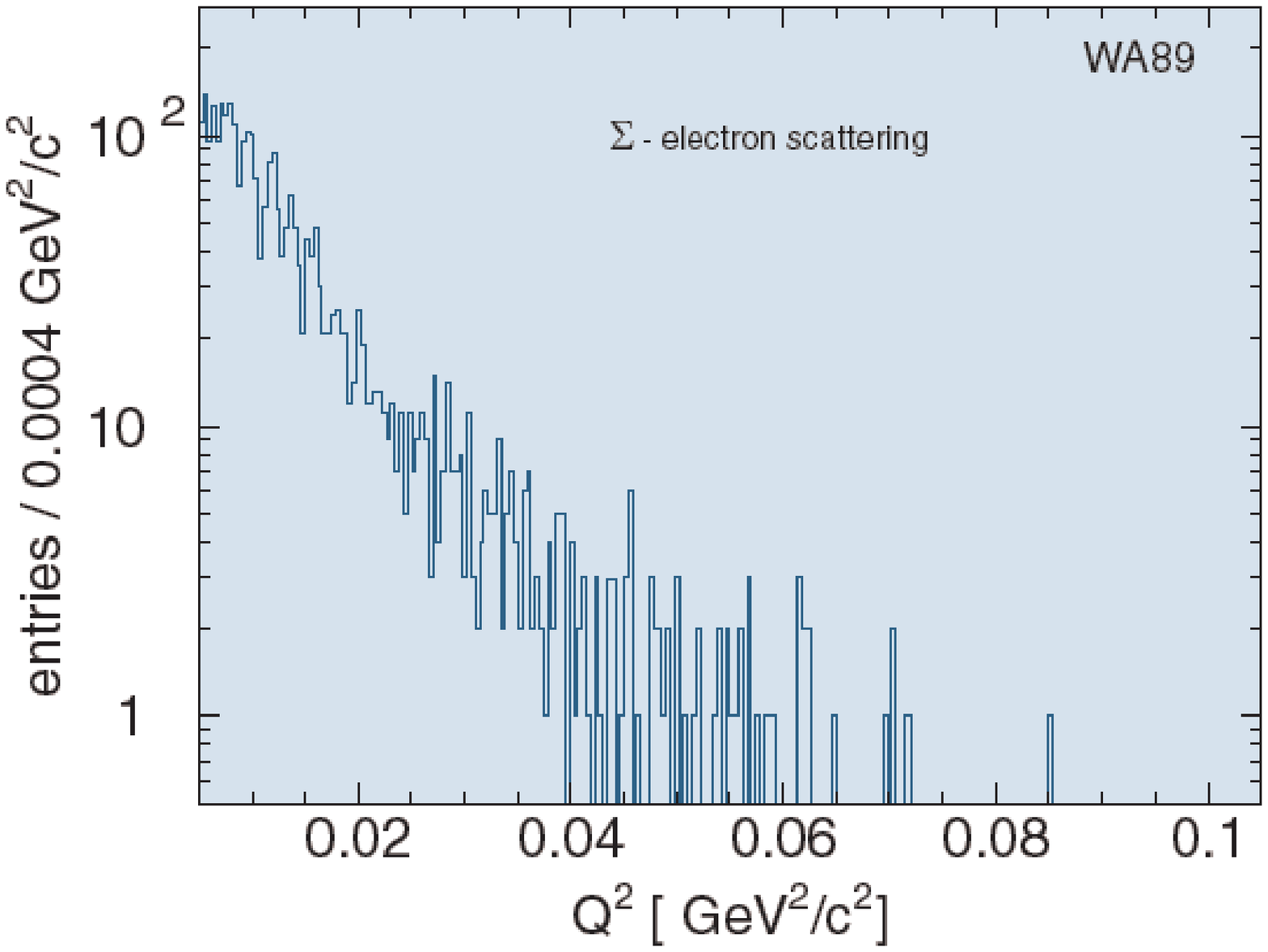,height=5cm,width=6cm}
\label{electron_scattering}

\hspace{5.8cm}
\begin{minipage}[h]{6cm}
\caption{First measurement on the differential distribution for
$\Sigma^-$-e$^-$-scattering in a hyperon beam at CERN.}
\end{minipage}
\end{figure}

\vspace{.2cm}

\subsection{Spectroscopy of strange and charmed baryons}
Before moving to heavy baryons, we shall mention new results on
the field of strange baryons.

\vspace{-0.5cm}
\begin{figure}[h]
\hspace{.5cm}
\psfig{figure=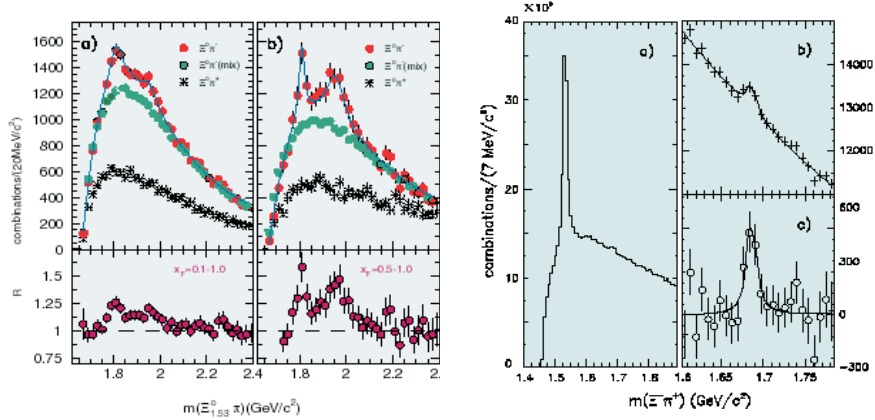
,height=6cm,width=12cm}


\caption{Largest data sample on $\Xi^*$ resonances observed in the CERN hyperon beam experiment.
Left: $\Xi^*\pi$ invariant mass distributions showing the $\Xi$(1820), $\Xi$(1955).
Right: $\Xi\pi$ invariant mass distributions
exhibiting a signal for the $\Xi^*$(1530) and $\Xi^*$(1690)}
\label{wa89_xistar_2}
\end{figure}

\clearpage

\newpage

\begin{figure}[ht]
\psfig{figure=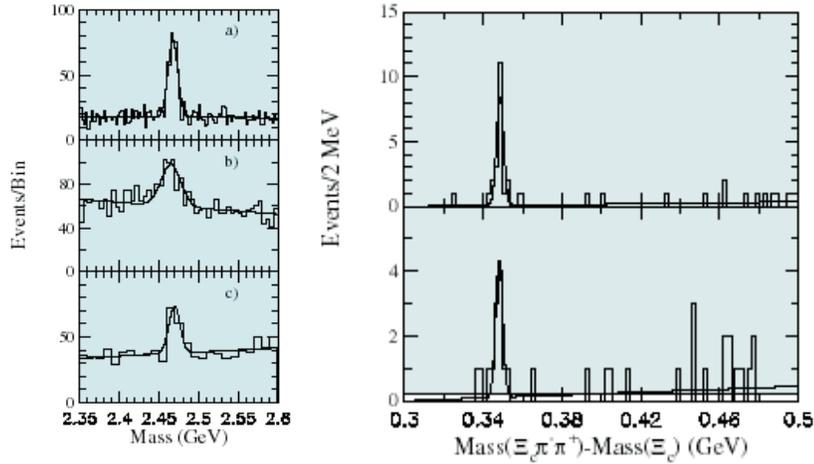,height=6.5cm,width=11cm}
\caption{Left histograms: Sum of all $\Xi_c$-daughter baryons in
various decay channels. Right histograms: invariant mass
difference of $\Xi_c\pi\pi$ and $\Xi_c$, latter one for positive and neutral charge states
(upper and lower, respectively). Not shown are the
mass distributions for the intermediate $\Xi_c\pi$-states, showing a dominant signal
at the $\Xi_c^*$-mass.}
\label{cleo_new_xic}
\end{figure}

\noindent
The most recent results on
spectroscopy (and the only one shown at this conference) concern
the largest data sample on $\Xi^*$-resonances observed (see fig.6), namely the
$\Xi$(1690), $\Xi$(1820), $\Xi$(1955) decaying into $\Xi\pi$ and
$\Xi^*$(1530)$\pi$, respectively\cite{WA89_hyperon}.
For the $\Xi$(1690)
its the firm confirmation of its existence,
for the other two the first observation of this decay channel. The
latter ones are produced predominantly in forward direction
($x_F\geq$0.5) indicating a diffractive production process.
\par
Newly observed states in the sector of charmed strange baryons
come from CLEO\cite{CLEO_excited}. They have identified for the first time an
orbitally  excited $\Xi_c^* (3/2^-)$, decaying into $\Xi_c^*
(3/2^+)$+$\pi$ (see fig.7). Although showing an excitation energy of about 350
MeV the state is still narrow. The explanation for its width
together with the spin-parity assignment is shown in fig.8.
Here we have separated two classes of states, those with l=0 and
l=1, the angular momentum of the light quark pair. Spin parity of
the light quark pair is indicated on the figure together with the
spin-parity of the total baryonic system. We assume that the spin
of the heavy quark does not participate in the interaction (full
separation of {\it light} and {\it heavy} system). Considering the
light systems, the $3/2^-$ state can go via S-wave to the $3/2^+$
state and via D-wave to the symmetric $1/2^+$ states.

\noindent
\begin{minipage}[h]{5cm}
Similarly, the $1/2^-$
state can go via D-wave to the $3/2^+$ and via S-wave to the
symmetric $1/2^+$ state. In both cases the decay to the anti-symmetric
ground state could be
forbidden by parity conservation in the light diquark. Thus, the $3/2^-$
state is expected to be narrow and decay predominantly to the
$3/2^+$-state, as observed. The $1/2^-$-state will preferably
decay into the symmetric ground state difficult to observe.
These arguments follow the line already employed for the
$\Lambda_c$-system.
\end{minipage}

\vspace{-2.8cm}

\begin{figure}[ht]
\hspace{5.8cm}
\psfig{figure=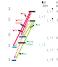,height=.8cm,width=.7cm}

\hspace{6cm}
\begin{minipage}[htb]{5.5cm}
\caption{Decay scheme for $\Xi_c$-resonances in HQET assuming
decoupling of light and heavy quarks}
\end{minipage}
\label{hqet}
\end{figure}

\par
Fig.9
shows the excitation spectrum for the
strange and heavy baryons. The zero-point of the energy scale is
matched to the lowest lying baryon-states of the flavor family. As
is expected from quark models the spin symmetry of a light quark
pair governing the HF-interaction leads

\noindent
\begin{minipage}[h]{5.5cm}
to an increased $\Lambda$ and $\Sigma$
splitting states with increasing {\it heavy} quark mass ($\Lambda -\Sigma$ vs. $\Lambda_c -\Sigma_c$, which is
compensated if the mass of the {\it light} quark system is
increased ($\Lambda_c -\Sigma_c$ vs. $\Xi_c -\Xi_c'$).
The $1/2^+$-$3/2^+$ splitting decreases with
the mass of the {\it heavy} quark but increases with the mass of
the {\it light} system ($\Sigma$ vs $\Xi$, $\Sigma_c$ vs
$\Xi_c$).
The excitation energy of the $3/2^-$ states slightly increases
with the mass of the {\it light}-system ($\Lambda_c$ vs $\Xi_c$).
As the light and strange quark pairs change role for the $\Sigma$
and $\Xi$ (and thus the definition of l and L) the situation is
different in the strange sector.
\par
Unfortunately, no published data yet
exist for the b-baryon sector.
\end{minipage}

\vspace{-7.5cm}

\begin{figure}[h]
\hspace{5.6cm}
\psfig{figure=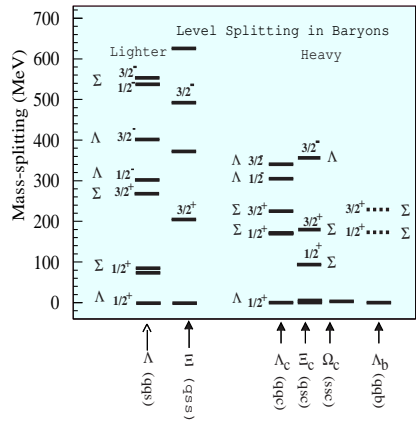,
height=5.8cm,width=5.8cm}


\hspace{6.cm}
\begin{minipage}[h]{5.5cm}
\caption{Summary of heavy baryon spectroscopy as of today. Denote
the zero-suppression of the mass scale for different flavor
compositions (s,c,b-quarks)}
\end{minipage}
\label{spectrum_1}
\end{figure}

\clearpage

\newpage

Previously reported results on the
$\Sigma_b$ and $\Sigma_b^*$ showed anomalously high spin splitting
which where not confirmed. However, the mass of the $\Lambda_b$
has now been measured by CDF\cite{CDF_1} in the decay channel
$\Lambda_b\rightarrow J/\Psi\Lambda$ to 5624$\pm$9 MeV/c$^2$.

\section{Death of heavy baryons - Decay Properties}

The interest in studying decays of heavy baryons is again related
to the static colour source and the large mass of the decaying
quark. The large phase space available in the quark decay (Q-value
$\gg m_{\pi}$) results in little influence of particular selection
rules like isospin in the strange baryon sector. In addition, the
decay of the baryon may be considered as a two-step process: decay
of the heavy quark (at first order independent of its surrounding)
and subsequent fragmentation of the resulting system. However,
this simple picture has proven wrong as is reflected in the large
variation of baryon lifetimes even for b-baryons. This has its
origin in the large effects of quark correlations which in turn
depend on the wave function overlap at the origin. While the free quark
decay rate is governed by phase space ($\sim m_Q^5$) and quark
mixing ($\sim V_{cs}$, $V_{cd}$ and $V_{bc}$) the influence of
the other quarks decreases with the decreasing localization overlap of heavy
and light quarks ($\Delta r\sim 1/m_Q^2$). The result is a modification of
branching fractions and decay asymmetries.

\subsection{Lifetimes of heavy baryons}

\noindent
\begin{minipage}[h]{5cm}
Fig.10 depicts
two possible processes (out of four) governing the heavy baryon
decay, namely the 'free quark decay' (or spectator diagram) and
the 'W-exchange' process. The first graph has a $m_Q^5$-mass
dependence, the second one $m_Q^3$.
Fig.11
depicts the measured life time pattern of heavy hadrons\cite{PDG}. The size of the
box indicate the present accuracy. The latest measurements in the
baryon sector have increased the accuracy for the $\Xi_c^+$ (E687) \cite{E687_life}
and $\Lambda_b$ (CDF\cite{CDF_LB} and LEP\cite{LEP_LB}).
\end{minipage}

\vspace{-8cm}

\begin{figure}[h]
\hspace{6cm}
\psfig{figure=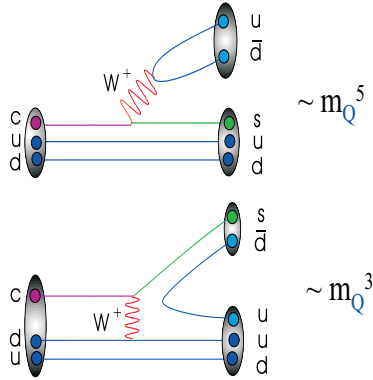,height=5cm,width=5cm}

\hspace{6.1cm}
\begin{minipage}[h]{6cm}
\caption{Two decay diagrams for the $\Xi_c^+$. The upper figure depicts the free
quark decay, the lower one W-exchange. Other decay diagrams exist, not shown here.}
\end{minipage}
\label{quark_correlations}
\end{figure}

\clearpage

\vspace{0.5cm}

\begin{figure}[h]

\hspace{5.5cm}
\psfig{figure=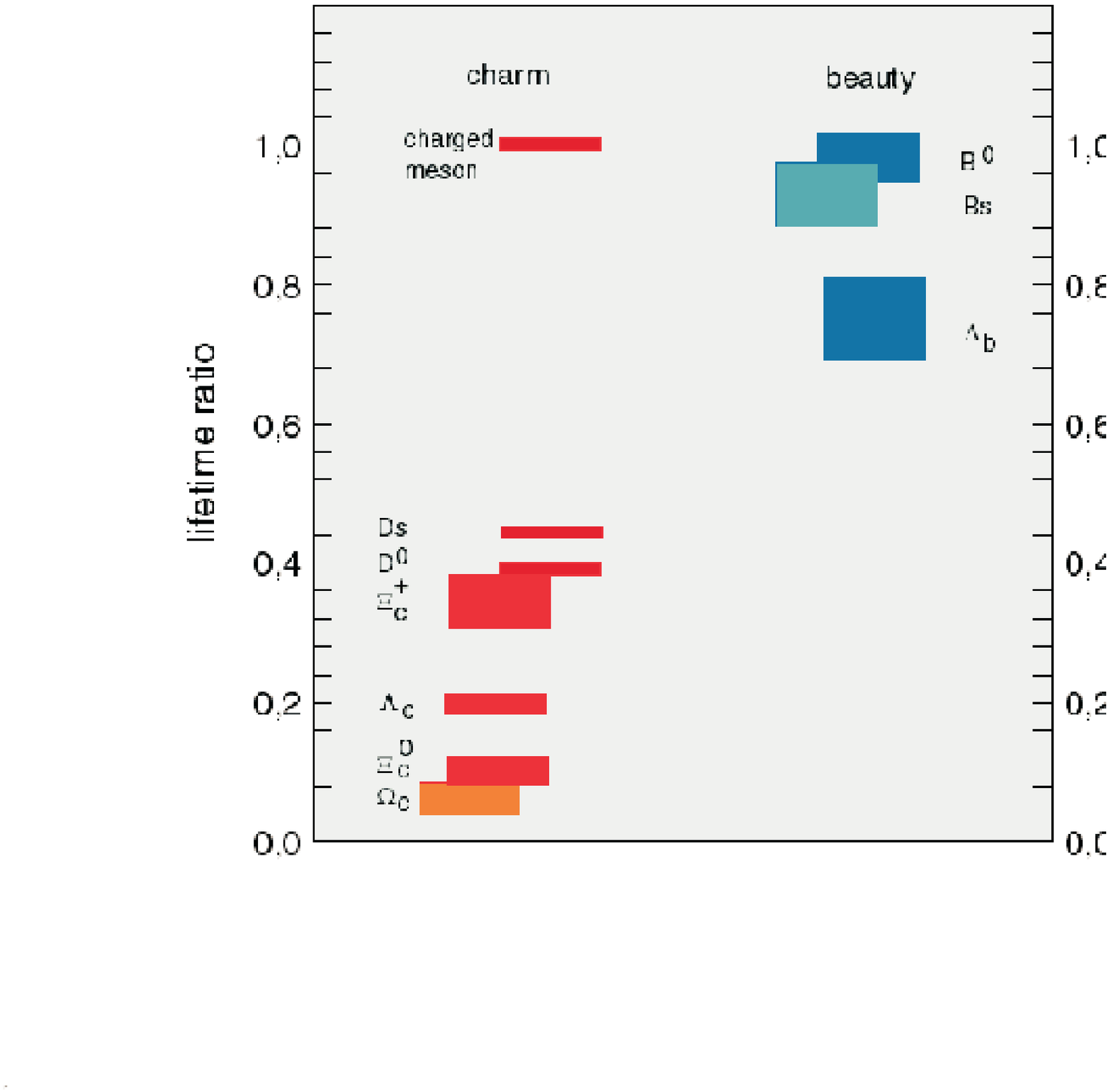,height=6cm,width=6cm}


\hspace{6.1cm}
\begin{minipage}[h]{5cm}
\caption{Measured lifetimes of heavy flavor hadrons. The vertical
size of a bar gives the error on the measured values.}
\end{minipage}
\label{lifetimes}
\end{figure}

\vspace{-7.7cm}

\noindent
\begin{minipage}[h]{5cm}
In particular, the low
lifetime of the $\Lambda_b$ has become more significant.
The central value for the world average is 1.24$\pm$0.08ps (0.74
of the B-meson lifetime) and can not be accommodated by the
present concepts and calculations.
\newline
Two conclusions may be drawn from this. The technique of Heavy
Quark Expansion is not working for the b-sector and thus the
quantitative agreement in the charm sector is pure coincidence
\cite{bigi}. Particular possibly hidden selection rules may alter
the decay width of some dominant decay channels \cite{lipkin}.
\end{minipage}



\subsection{Analysis of simple final states - Decay asymmetries}
While the description of multi-body decays, which constitutes a
large part of the observed total decay width of heavy baryons, is very
difficult to describe theoretically, simple final states (like
two-body decays or three-body semi-leptonic decays) may be calculated.
This is possible if we assume the factorization of the decay
process\cite{bjorken}:
\begin{itemize}
  \item separate W-emission and W-decay vertex (heavy quark transition)
  \item Only assume spectator diagram (W-emission) (see fig.10
  upper graph)
  \begin{itemize}
     \item W-decay into lepton pair (l$^{\pm}\nu$) $\rightarrow$ semi-leptonic
     decay
     \item W-decay in $\pi$ (or K-Cabibbo suppressed) $\rightarrow$ two-body
     non-leptonic decays
  \end{itemize}
\end{itemize}
then:
\begin{itemize}
  \item for very heavy baryons
  $\Gamma_{s.l.}$=$\Gamma_{s.l.}^{free quark}$ - branching ratio (b.r.) is indicator
  for importance of quark correlations to lifetime (see D$^{\pm}$
  with b.r. (semi-leptonic)$\sim$15\%)
  \item predict 2-body decay asymmetries for $\Lambda$-type
  baryons
    \begin{itemize}
        \item $\Lambda_c\rightarrow\Lambda\pi$ $\alpha$=-1
        \item $\Xi_c\rightarrow\Xi\pi$ $\alpha\leq$0 (W-exchange
        is important)
        \item HQET can make predictions (only 2 decay amplitudes
        may remain)
    \end{itemize}
\end{itemize}
\par
Typically, weak decay asymmetries are caused by the interference
of S- and P-waves in the decay of the mother baryon. This leads to
parity violation which is manifested in a polarization of the
daughter particles or in an asymmetry in their decay angle with
respect to the polarization of the mother baryon, both governed by
the analyzing power $\alpha$ of the decay. Thus, the analysis of
the daughter polarization can be used to study the analyzing power
in a specific decay channel (see fig. 12).
HQET predicts a strong correlation between s.l. decay asymmetries
and non-leptonic two-body decays of type $\Lambda_c\rightarrow\Lambda$ for
q$^2$=0, predicting $\alpha_{\Lambda_c}$=-1 (as observed by CLEO\cite{CLEO_LC}).
Recently, non factorizing diagrams have also been included in the
calculations leading to more realistic predictions of the decay
properties\cite{koerner}.

\vspace{-.5cm}

\begin{figure}[h]
\hspace{6cm}
\psfig{figure=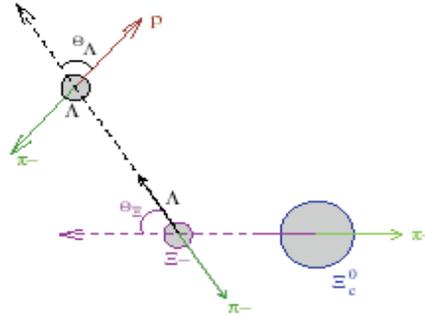,height=7cm,width=6cm}
\label{xic_asymmetry}

\vspace{-3cm}

\hspace{6.4cm}
\begin{minipage}[htb]{6cm}
\caption{Determination of the decay asymmetry of a $\Xi_c$, using
the decay chain with self-analyzing hyperons}
\end{minipage}
\end{figure}

\vspace{-5.5cm}

\noindent
\begin{minipage}[ht]{5cm}
Recently CLEO has made the first measurement on a decay asymmetry
of the $\Xi_c^0$ observing the triple cascade
$\Xi_c\rightarrow\Xi^-\pi^+$, $\Xi^-\rightarrow\Lambda\pi^-$,
$\Lambda\rightarrow p\pi$ \cite{CLEO_asymmetry}. Using the known decay
asymmetries for the hyperons the decay asymmetry of the 2-body
decay channel could be determined. The preliminary result is
$\alpha_{\Xi_c^0}$$=-0.56\pm 0.39$. This can be compared to the
analog decay $\Lambda_c\rightarrow\Lambda\pi^-$ with
$\alpha_{\Lambda_c^+}$=-0.94$^{+0.21}_{-0.06}$.
\end{minipage}

\begin{figure}[h]
\psfig{figure=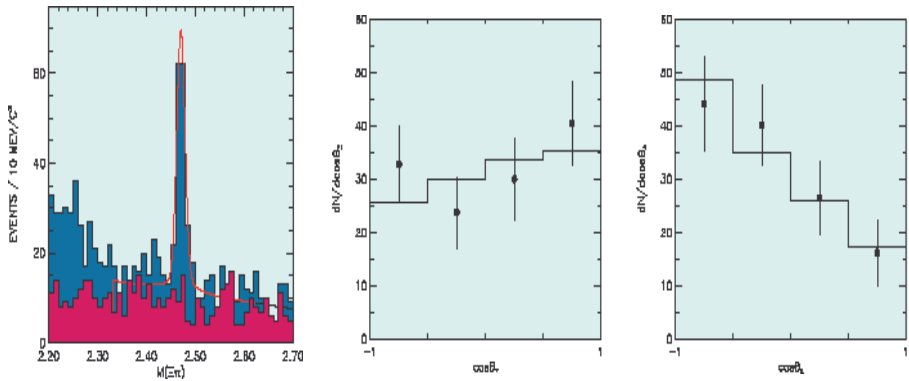,height=5cm,width=12cm}
\caption{Underlying signal for the two-body asymmetry analysis of
the $\Xi_c$ (left) and angular asymmetries observed (right two
plots). The angular distributions (data poins) are compared with MC-calculations
(histogram) based on the resulting asymmetry parameter.}
\label{cleo_asymmetry_results}
\end{figure}

Although statistically still poor this measurement prefers
predictions with negative sign of the analyzing power.

\section{Is Their Life After Death ?}
After having spent many decades on the study of heavy
(predominantly charmed) baryons we shall discuss the role of heavy
baryons in future experiments. We may distinguish two fields, the
search for new heavy baryonic systems and the use of heavy baryons
as tools to study other physics subjects.
\subsection{Future Cultures}
The investigation of new systems comprises ordinary baryons and
exotica. The next system which might be investigated are double
charmed baryons. The physics of these lies in the spectroscopy\cite{ccbaryon} and
the lifetime pattern observed for different flavor states.
\newline
While the ground state of such a system looks like a deuterium
atom with the two charm quarks taking the role of the nucleus, the
first excited levels may have molecular structure with {\it
vibrational} states of the two heavy quarks building the lowest
levels of the excitation spectrum. The lifetime pattern of the
different possible flavor states should enable us to unravel the
different long distance contributions mentioned above as only one
additional process adds to the spectator diagram, a different one
for each baryon type.
\par
However, we will face experimental difficulties. The cross
sections at $e^+e^-$ machines are very low and still low in fixed
target production. High beam intensities are required (see e.g.
COMPASS\cite{COMPASS}, BTEV\cite{BTEV}) and very high energy may be necessary to boost the
cross section for double charm production (see B$_c$-production at
the Tevatron). In addition to the production the detection is non
straight forward. Four decay vertices from the 4 charm quarks
produced in the interaction have to be disentangled on very short
distances.

\subsection{Heavy Baryons as a tool - the case of polarization}
We may use the production properties of heavy baryons as a tool to
measure other physics parameters. The analyzing power in two-body
and semi-leptonic decays can be used to measure their production
polarization. Those can be used to measure:

\begin{itemize}
  \item quark polarization in Z-decays\\
  The standard model gives predictions of large polarization of
  the b-quarks. This can be used to measure the polarized
  fragmentation function or, in turn test the b-quark polarization
  in some specific kinematic region\cite{aleph_polarization}.
  First results have already be obtained for the $\Lambda_b$
  polarization showing a surprisingly low value of
  $\wp_{\Lambda_b}$=-23$^{+24}_{-20}$(stat.)$^{+8}_{-7}$(syst.)\%.
  \item polarization of gluons in polarized nucleons\\
  The production of polarized $\Lambda_c$ has already been discussed
  within
  the COMPASS\cite{COMPASS} experiment as an alternate tool to measure the spin
  polarization of gluons. Using deep inelastic muon scattering
  c-quarks are produced by the photon-gluon fusion process.
  Baryons are subsequently formed
  in their fragmentation. Owing to the production
  process the polarization of the c-quarks will follow the gluon
  polarization in the nucleon. If we were to follow SU(3) w.f. the
  c-quark determines the spin-direction of the $\Lambda_c$. We may
  thus use two-body decays with large analyzing power to determine
  the $\Lambda_c$-polarization in the standard asymmetry
  measurements using the polarized fragmentation functions derived e.g. from LEP.
  Knowing the gluon polarization in the nucleon we could reverse the
  argument and determine the c-quark polarization in a polarized
  $\Lambda_c$.
\end{itemize}

\subsection{Other fields of interest}
We may briefly enumerate a number of other interesting
measurements still to be done in the sector of heavy baryons.
\begin{itemize}
  \item precise knowledge of form factors in semi-leptonic decays
  \item b-baryon spectroscopy
  \item spectrum of excited c-baryons (are they still narrow ?)
  \item magnetic moments (needs very high energy beams(e.g. 8 TeV
  extracted beams), high intensity, observation of spin precession
  in e.g. crystal)
\end{itemize}
Last not least we shall remember, that the search of doubly
strange dibaryons has not yet revealed any sign of their existence
(see experiments at BNL \cite{BNL}, KEK \cite{KEK} and CERN
\cite{WA89_H}) and that hybrid baryons are still discussed as
possible candidates for exotic objects.
\section{Conclusion}
We have given a survey of the different physics surrounding the
study of heavy baryons.
\newline
It has been shown that
hadro-production reveals interesting features not found in simple
models of quark fragmentation and that total charm cross sections
still show large systematic uncertainty both from theory and
experiment.
\newline
The study of static properties of heavier baryons is being pursued
and first measurements of the charge radius of a strange baryon
has been reported. The spectroscopy of charmed baryons has seen
further discoveries in orbitally excited states supporting the
scheme of decoupling of heavy and lighter quark degrees of
freedom.
\newline
The lifetime puzzle in the b-quark sector persists. Still the
precision in baryon lifetimes is not yet high enough to definitely
challenge existing model calculations. Big progress is also at
the horizon in the field of decay asymmetries, a subject not
understood in the sector of lighter systems. Last not least it
should be mentioned, that most of the recent data stem from
photo-production in ether photon-beams or $e^+e^-$-colliders which
offer clean environment even for the study of low cross section
processes.


\begin{thebibliography}{99}

\bibitem{ridolfi}S. Frixione et al., hep-ph/9702287
\bibitem{WA89_charm}Y. Alexandrov et al., submitted and accepted to {\it Eur. Phys. J.}
C, hep-ex/9803021
\bibitem{SELEX}J. Russ, CMU-HEP 98-07, presentation at ICHEP98, Vancouver, 1998
\bibitem{FOCUS}P. Sheldon, presented at ICHEP98, Vancouver, 1998
\bibitem{aleph_baryon_counting}R. Barate et al., {\it Eur. Phys.
J.} {\bf C5}, 205, (1998)
\bibitem{Isgur}see e.g. N. Isgur and G. Karl, \Journal{\PRD}{19}{2653}{1979}
or \Journal{\PRD}{18}{4187}{1978}
\bibitem{WA89_sigel}Y. Alexandrov et al.,submitted and accepted to {\it Eur. Phys. J.} {\bf C}, (1998)
\bibitem{SELEX_sigel}I. Eschrich, contribution to this conference,
hep-ex/9811003
\bibitem{WA89_hyperon}Y. Alexandrov et al., {\it Eur. Phys. J.} {\bf C5}, 621, (1998)
\bibitem{CLEO_excited}CLEO collaboration, CLEO/Conf 98-10, presented at
ICHEP98, Vancouver, 1998
\bibitem{CDF_1}F. Abe et al., \Journal{\PRL}{77}{1439}{1996}
\bibitem{PDG}Particle Data Group, {\it Eur. Phys. J.} {\bf C3}, 1998
\bibitem{E687_life}P.L.Frabetti et al., \Journal{\PLB}{427}{211}{1998}
\bibitem{CDF_LB}F. Abe et al., \Journal{\PRD}{55}{1142}{1997}
\bibitem{LEP_LB}see e.g. R. Alkers et al., \Journal{\PLB}{353}{402}{1995}
\bibitem{bigi}I. Bigi, hep-ph/9612293 and I. Bigi, M. Shifman, N.
Uraltsev, {\it Ann. Rev. Nucl. Sci.} 47, 591, 1997
\bibitem{lipkin}A. Datta, H. Lipkin and P.J. O'Donnell,
hep-ph/9809294
\bibitem{bjorken}J.D. Bjorken, \Journal{\PRD}{40}{1513}{1989}
\bibitem{CLEO_LC}G. Crawford et al., \Journal{\PRL}{75}{624}{1995}
and M. Bishai et al., \Journal{\PLB}{350}{256}{1995}
\bibitem{koerner}M.A.Ivanov et al., \Journal{\PRD}{57}{5632}{1998}
\bibitem{CLEO_asymmetry}I. Shipsey, presented at ICHEP98,
Vancouver, 1998,CLEO/Conf 98-16
\bibitem{aleph_polarization}D. Buskulic et al., \Journal{\PLB}{374}{319}{1996} and
\Journal{\PLB}{365}{183}{1996}
\bibitem{ccbaryon}S. Fleck and J.M. Richard, {\it Prog. Theor.
Phys.}, 82, 760, 1989
\bibitem{COMPASS}The COMPASS Collaboration,
CERN-SPSLC-96-14, 1996
\bibitem{BTEV}BTEV-collaboration, Nucl. Instr. Meth. {\bf A408}, 146, 1998
\bibitem{BNL}K. Yamamoto et al., \Journal{\NPA}{639}{371}{1998}
and references therein
\bibitem{KEK}J.K. Ahn et al., KEK-Preprint, 98-24 and
\Journal{NPA}{639}{21c}{1998}, \Journal{\PRL}{65}{1729}{1990}
\bibitem{WA89_H}M. Godbersen, Doctoral thesis, Heidelberg, 1991
and M. Beck, Doctoral thesis, Heidelberg, 1996
\end{thebibliography}
\end{document}